%% file: main_noVT_arXiv.tex
\documentclass{llncs}
\usepackage{hyperref}
\usepackage[x11names]{xcolor}
\hypersetup{
colorlinks=true,
citecolor=blue
}
\usepackage{orcidlink}
\usepackage{amssymb,amsmath}
\usepackage{mathtools}
\usepackage{xspace}
\usepackage{graphicx}
\usepackage{comment,multirow} 
\usepackage{bm,mathcomp,tikz,tabularx}
\usepackage{tikz}
\usepackage[figure,tworuled,linesnumbered,vlined]{algorithm2e} 

\usepackage[a4paper, margin=1.3in]{geometry}

\input{commands}

\title{Gathering in Non-Vertex-Transitive Graphs  Under Round Robin}

\author{}
\institute{}
 
\author{Serafino Cicerone\inst{1} \and Alessia Di Fonso\inst{1} \and Gabriele Di Stefano\inst{1} \and \\ Alfredo Navarra\inst{2}}
 \institute{Dipartimento di Ingegneria e Scienze dell'Informazione e Matematica, \\
         Universit\`a degli Studi dell'Aquila, Via Vetoio I-67100 
         L'Aquila, Italy. \\
 \email{\{serafino.cicerone, alessia.difonso, gabriele.distefano\}@univaq.it}
 \and
 Dipartimento di Matematica e Informatica, 
         Universit\`a degli Studi di Perugia,\\ Via Vanvitelli I-06123 
         Perugia, Italy.
 \email{alfredo.navarra@unipg.it}
}

\begin{document}

\maketitle

\begin{abstract}
The {\gath} problem for a swarm of robots asks for a distributed algorithm that brings such entities to a common place, not known in advance. We consider the well-known $\mathcal{OBLOT}$ model with robots constrained to move along the edges of a graph, hence gathering in one vertex, eventually. Despite the classical setting under which the problem has been usually approached, we consider the `hostile' case where: i) the initial configuration may contain \emph{multiplicities}, i.e. more than one robot may occupy the same vertex; ii) robots cannot detect multiplicities. As a scheduler for robots activation, we consider the ‘favorable’ \emph{round-robin} case, where robots are activated one at a time.

Our objective is to achieve a complete characterization of the problem in the broad context of \emph{non-vertex-transitive} graphs,  i.e., graphs where the vertices are partitioned into at least two different classes of equivalence. We provide a resolution algorithm for any configuration of robots moving on such graphs along with its correctness. 
Furthermore, we analyze its time complexity.
\end{abstract}

% ==================================================================

% ------------------------------------
% INTRODUCTION
% ------------------------------------
\section{Introduction}
A particularly prominent model 
for swarm robotics %
from a theoretical perspective is the $\mathcal{OBLOT}$ model~\cite{FPS-macbook19,FPS12}, which characterizes robots with weak computational power. In this framework, robots operate according to repeated {\tt Look-Compute-Move} cycles: during each cycle, a robot observes its surroundings ({\tt Look}), computes its next move based on a deterministic algorithm ({\tt Compute}), and then proceeds to the selected destination ({\tt Move}).

One of the most studied problems in this setting is the so-called {\gath} problem. Robots are required to converge to a common, unspecified place where they eventually stop moving. For robots moving in the Euclidean plane, the {\gath} problem has been fully characterized. In fact, it is always solvable for synchronous robots, whereas in the asynchronous case it is unsolvable when just two robots are considered. %This last case with just two robots is usually referred to as the \emph{rendezvous} problem. 

When robots are constrained to move along the edges of a graph, the situation is less clear. Apart from some impossibility results or basic conditions that guarantee the resolution of the {\gath} problem
provided in \cite{CDN18-book,CDN20a,DN17a}, most of the literature usually focuses on specific topologies.
Studied topologies are: 
Trees \cite{DDKN12,DDN13}, Regular Bipartite graphs~\cite{GP13}, Finite Grids \cite{DDKN12}, 
Infinite Grids \cite{DN17}, 
Tori \cite{KLOTW21}, Oriented Hypercubes \cite{BKAS18}, Complete graphs~\cite{CDN19c,CDN20a}, 
Complete Bipartite graphs \cite{CDN19c,CDN20a}, Butterflies \cite{CDDN24}, 
and Rings \cite{BPT16,DDN14,DDN11,DDNNS15,DNN17,DDFN18,DN17a,IIKO13,KKN10,KMP08,NP25,OT12}.	

The most of such topologies are very symmetric, i.e., the vertices can be partitioned into a few classes of equivalence. %Since robots have few topological properties to exploit, the design of a resolution algorithm becomes more challenging.
%In particular, \emph{vertex-transitive} graphs, i.e., graphs where all the vertices are topologically equivalent, result to be very difficult to approach. Examples of such graphs are complete graphs, complete bipartite graphs $K_{n,n}$, rings, hypercubes or infinite grids.
%On the other hand, in non-vertex-transitive graphs, the topology may provide some help to accomplish the {\gath}, but this is not always that simple, see, e.g. Butterflies \cite{CDDN24}. 
%
Another crucial aspect of the cited works, applicable to both Euclidean and graphs, is the time scheduling under which the robots operate. Notably, the \emph{asynchronous} scheduler, where robots are activated independently of one another, often presents the greatest challenges. However, there are cases where asynchrony makes the problem unsolvable, whereas the \emph{synchronous} setting allows the development of non-trivial strategies, see, e.g.~\cite{CDN20a}.
In any case, there are two very common assumptions in the literature: 
\begin{description}
    \item[$A_1$:] robots are endowed with some kind of {\emph{multiplicity detection}}. With this, robots are able to recognize whether a vertex contains a {\emph{multiplicity}}, i.e., if two or more robots are located at the same vertex (not necessarily the exact number); 
    \item[$A_2$:] the \emph{initial} configuration, i.e., the first configuration ever seen by any robot,  does not contain multiplicities.     
\end{description}
In a recent work~\cite{FW24}, the version of the {\gath} problem that assumes $A_2$ is denoted as {\dgath}, while {\gath} refers to the case without that assumption. From now on, we keep this distinction. As a time scheduler, we consider the {\emph{Round Robin}} (\rr). This is a specific type of synchronous scheduler, where $k$ robots are activated one at a time in a fair sequence. That is, in the first $k$ {\tt Look-Compute-Move} cycles, all robots are activated. This constitutes the first \emph{epoch}. Then, another epoch starts with the same order of activations fixed by an ideal adversary and unknown to the robots and repeating forever. %In the general case, instead, the order of activation of the robots can be changed by 
%an ideal adversarial scheduler in each epoch. This might be referred to as the \emph{aperiodic}-\rr.
Certainly, dealing with {\rr} may seem much easier compared to other schedulers, especially the asynchronous one. Indeed, all the symmetries admitted by a given configuration are inherently broken each time by the single activated robot. 
However, the {\gath} problem becomes considerably more complicated without assumptions $A_1$ and $A_2$.
%However, when dealing with the {\gath} problem without assumptions $A_1$ and $A_2$ highly complicates the problem.
In~\cite{FW24}, the {\gath} problem has been fully characterized when robots freely move on the Euclidean plane: it is impossible for $k=3$ robots but possible for any $k\ge 4$. 
%
%It is worth mentioning that the more generic {\emph{sequential}} scheduler, which requires only to activate one robot 
%at a time with some eventual fairness, has been approached in \cite{FNPPS2024} to solve the {\emph{Universal Pattern Formation}} ({\tt UPF}) problem. In {\tt UPF}, the requirement is to move robots so as to form a given pattern in the Euclidean plane, without assumptions $A_1$ and $A_2$.
%Surprisingly, the only pattern that was not possible to guarantee is the \emph{point}, i.e., under such a scheduler {\gath} is unsolvable. Concerning robots moving on graphs, in~\cite{NP25}, both {\dgath}  and {\gath} problems have been addressed for robots moving on rings under periodic and general {\rr} schedulers. In particular, in the context of rings, a complete characterization of the {\gath} problem under \rr\ is given.

\block{Our results}
In this paper, we continue the investigation of the {\gath} problem under the newly proposed \rr\ scheduler, without relying on assumptions $A_1$ and $A_2$, and considering general graphs. In particular, our goal is to achieve a complete characterization of the problem in the broad context of non-vertex-transitive graphs. For these graphs, we demonstrate that the concepts of orbits\footnote{Roughly speaking, an orbit is a maximal equivalence class of vertices, i.e., a maximal set of indistinguishable vertices.} and graph canonization\footnote{A method for finding a labeling of a graph $G$ such that every graph isomorphic to $G$ has the same labeling as $G$.} serve as powerful analytical tools. In particular, we present a simple, time-optimal algorithm, $\algoterm$, designed for graphs that contain a terminal orbit — a novel graph-theoretic concept introduced in this work and linked to other graph-theoretic parameters.
For graphs that lack terminal orbits, we identify structural properties that require the development of a more intricate gathering algorithm, $\algonoterm$. 
Nevertheless, it is a linear factor away, in the graph's size, from being time-optimal in terms of epochs.

%\block{Outline}
%In the next section, we recall the $\mathcal{OBLOT}$ model. In Section~\ref{sec:prob}, the {\gath} problem is formulated along with some useful notation. Section~\ref{sec:non-vertex} contains all the results obtained for non-vertex-transitive graphs. In particular, we first provide a general overview of the amplitude of this family of graphs, then we fully characterize the case by providing a general resolution algorithm along with its correctness.  Finally, Section~\ref{sec:concl} contains concluding remarks and future work suggestions.

% ------------------------------------
% MODEL
% ------------------------------------
\section{Model}\label{sec:model}
We consider the standard  $\mathcal {OBLOT}$ model of 
distributed systems of autonomous mobile robots. In $\mathcal {OBLOT}$, the system is composed of a set 
 $\mathcal{R} = \{r_1, r_2, \dots, r_k\}$ of $k\ge 2$ computational {\emph robots} that  operate on a graph  $G$.
 Each vertex of $G$ is initially empty, occupied by one robot, or occupied by more than one robot (i.e., a \emph{multiplicity}; recall that we are not using assumptions $A_1$ and $A_2$ as described in the Introduction). Robots can be characterized according to many different settings. In particular, they have the following basic properties: 
  \begin{itemize}
      \item \textbf{Anonymous:} they have no unique identifiers;
      \item \textbf{Autonomous:} they operate without a centralized control;
      \item \textbf{Dimensionless:} they are viewed as points, i.e., they have no volume nor occupancy restraints;
      \item \textbf{Disoriented:} they have no common sense of orientation;
      \item \textbf{Oblivious:} they have no memory of past events;
      \item \textbf{Homogeneous:} they all execute the same deterministic algorithm with no type of randomization admitted;
      \item \textbf{Silent:} they have no means of direct communication.
  \end{itemize}
Each robot in the system has sensory capabilities, allowing it to determine the location of
other robots in the graph, relative to its location. Each robot refers to a {\tt Local Reference System}
({\tt LRS}) that might differ from robot to robot. %Each robot follows an identical algorithm that is pre-programmed into the robot. 
Each robot has a specific behavior described according to the sequence of the following four states: {\tt Wait}, {\tt Look}, {\tt Compute}, and {\tt Move}. Such a sequence defines the computational activation cycle (or simply a cycle) of a robot. More in detail:

\begin{enumerate}
    \item {\tt Wait:} the robot is in an idle state and cannot remain as such indefinitely;
    \item {\tt Look:} the robot obtains a global snapshot of the system containing the positions of the other robots with respect to its {\tt LRS}, by activating its sensors. Each robot is seen as a point in the graph occupying a vertex;
    \item {\tt Compute:} the robot executes a local computation according to a deterministic algorithm $\mathcal{A}$ (we also say that the robot executes 
    $\mathcal{A}$). This algorithm is the same for all the robots, and its result is the destination of the movement of the robot. Such a destination is either 
    the vertex where the robot is already located, or a neighboring vertex at one hop distance (i.e., only one edge per move can be traversed); 
    \item {\tt Move:} if the computed destination is a neighboring vertex $v$, the robot moves to $v$; otherwise, it executes a {\emph{nil}} 
    movement (i.e., it does not move).
\end{enumerate}

In the literature, the computational cycle is simply referred to as {\tt Look-Compute-Move} ({\tt LCM}) cycle, because when a robot is in the {\tt Wait} state, we say that it is 
\emph{inactive}. Thus, the {\tt LCM} cycle only refers to the \emph{active} states of a 
robot. 
It is also important to notice that since the robots are oblivious, without 
memory of past events, every decision they make during the {\tt Compute} phase is 
based on what they can determine during the current {\tt LCM} cycle. In 
particular, during the {\tt Look} phase, the robots take a global snapshot of the system and 
they use it to elaborate the information, building what is called the \emph{view} of the 
robot. 
Regarding the {\tt Move} phase of the robots, the movements executed are always considered to be instantaneous. Thus, the robots are only able to perceive the 
other robots positioned on the vertices of the graph, never while moving. Regarding the 
position of a robot on a vertex, two or more robots may be located on 
the same vertex, i.e., they constitute a multiplicity.

A relevant feature that greatly affects the computational power of the robots is the \emph{time scheduler}. 
In this work, we consider the standard Round Robin (\rr): 
\begin{itemize}
    \item 
   Robots are activated one at time. The time during which a robot is active is called a \emph{round}.
   If there are $k$ robots, then from round $1$ to round $k$, all the robots are activated. In the subsequent $k$ rounds,  all the robots are again activated in the same order. Each of those intervals of $k$ rounds is called an \emph{epoch}. The order in which robots are activated is decided at the beginning by an adversarial scheduler.\footnote{Actually, our strategies also work if the order of the activations is changed by the adversary at each epoch.}
\end{itemize}

% ------------------------------------
% PROBLEM
% ------------------------------------
\section{Problem formulation and preliminary observations}\label{sec:prob}

The topology where robots are placed on is represented by a simple and connected graph $G=(V,E)$, with finite vertex set $V(G)=V$ and finite edge set $E(G)=E$.
%(the only exception is in Section~\ref{sub:grids}, where we deal with infinite grids). 
The cardinality of $V$ is represented as $|V|$ or $n$. $G[X]$ denotes the subgraph of $G$ induced by a subset of vertices $X\subset V$. We denote by $diam(G)$ the diameter of $G$, that is, the maximum distance between any pair of vertices of the graph. For each vertex $v\in V$, $N(v)$ is the set of neighboring vertices of $v$ and $N[v] = N(v)\cup \{v\}$. Two vertices $u$ and $v$ are \emph{false twins} if $N(u)=N(v)$ and \emph{true twins} if $N[u]=N[v]$. A vertex $v$ is \emph{pendant} in $G$ if $|N(v)|=1$.

A function $\lambda: V\to \{0,1\}$  
%from the set of vertices to the set $\{0,1\}$ 
indicates to the robots whether a vertex of $G$ is empty or occupied. Note that more than one robot may occupy the same vertex, but robots cannot perceive such information. We call $C=(G,\lambda)$ a \emph{configuration} -- from the robots' perspective, whenever the actual number of robots is bounded and greater than zero.
%
%A vertex $v\in V$ such that $\lambda(v)> 0$ is said \emph{occupied}, \emph{unoccupied} otherwise. 
A subset $V'\subseteq V$ is said \emph{occupied} if at least one of its elements is occupied,  \emph{unoccupied} otherwise.
%
%A configuration is \emph{initial} if each robot lies on a different vertex (i.e., $\lambda(v)\leq 1$ for each $v\in V$). 
We denote by $\Delta(C)$ the maximum distance among any pair of vertices that are occupied in $C$, and by $occ(C)=\sum_{v\in G} \lambda(v)$ the number of vertices that are occupied in $C$.

A configuration $C$ is \emph{final} if all the robots are on a single vertex (i.e., $\exists u\in V: \lambda(u)=1$ and $\lambda(v)=0,~\forall v\in V \setminus \{u\}$), i.e. $\Delta(C)= 0$.
Any configuration $C$ that is not final can be \emph{initial}.
The gathering problem can be formally defined as the problem of transforming an initial configuration into a final one. Throughout the paper, we assume that each initial configuration is composed of at least two robots occupying at least two vertices (otherwise, the problem is trivially solved). A \emph{gathering algorithm} for this problem is a deterministic distributed algorithm that brings the robots in the system to a final configuration in a finite number of \LCM-cycles from any given initial configuration, regardless of the adversary. %
Formally, an algorithm $\A$ solves the gathering problem for an \emph{initial configuration} $C$ if,  for any execution $\Ex : C=C(0),C(1),C(2),\ldots$ of $\A$, there exists a time instant $t>0$ such that $C(t)$%
\footnote{$C(t)$ represents the configuration $C$ as observed at time instant $t$.}
is final and no robots move after $t$, i.e., $C(t') = C(t)$ holds for all $t'\ge t$. 
Given an initial configuration $C=(G,\lambda)$, it is worth remarking that many different placements of robots correspond to $C$ due to possible multiplicities. If there exists a gathering algorithm for all the placements of robots corresponding to $C$, we say that $C$ is \emph{gatherable}, otherwise we say that $C$ is \emph{ungatherable}.  
With respect to the number of epochs required to accomplish the gathering, we can state the following:

\begin{lemma}\label{lem:lb}
Given an initial configuration $C=(G,\lambda)$, any resolution algorithm for {\gath} on $C$ under \rr\ requires $\Omega(\Delta(C))$ epochs. 
\end{lemma}
\begin{proof}
Let $r$ and $r'$ be two robots occupying two vertices that determine $\Delta(C)$. In order to solve the {\gath} problem, $r$ and $r'$ should meet, eventually. The fastest way to do it is that they move toward each other along a shortest path. This requires $\Delta(C)/2$ moves for each robot and hence, due to the \rr\ scheduler, $\Delta(C)/2$ epochs.\qed
%As the maximum distance among robots in $C$ is at most $diam(G)$, by the generality of $C$, the claim holds.\qed
\end{proof}

Note that, as we are dealing with  finite graphs, the above lemma can be stated also with respect to $\Omega(diam(G))$ as it is easy to provide a configuration $C$ with $\Delta(C)=diam(G)$.

%\smallskip
We now introduce a tool that will play a fundamental role in describing a solution for the problem at hand.  In particular, we recall from~\cite{CDN20a} and~\cite{DN17a} the notions of automorphism and orbit for general graphs.

\subsection{Graph automorphisms and Orbits} 
Two undirected graphs $G=(V,E)$ and $G'=(V',E')$ are \emph{isomorphic} if there is a bijection $\varphi$ from $V$ to $V'$ such that $\{u,v\} \in E$ if and only if $\{\varphi(u),\varphi(v)\} \in E'$. An \emph{automorphism} on a graph $G$ is an isomorphism from $G$ to itself. The set of automorphisms of a given graph, under the composition operation, forms the automorphism group of the graph, denoted by $\Aut{G}$. Two vertices $u$ and $v$ of $G$ are \emph{equivalent vertices} if there exists an automorphism $\varphi\in \Aut{G}$ such that $\varphi(u)=v$. The equivalence classes of the vertices of $G$ under the action of the automorphisms are called vertex \emph{orbits}. The partition of the vertex set of $G$ consisting of the orbits by $G$
is called \emph{orbit partition} of $G$ and is denoted by $\orbs{O}_G$. It is easy to observe that when $G$ contains only one orbit, then it is \emph{vertex-transitive}, otherwise it is \emph{non-vertex-transitive}. We say that two orbits $O$ and $O'$ 
%(not necessarily distinct) 
are \emph{adjacent} if and only if there is an edge $(x,y)$ in $G$ such that $x\in O$ and $y \in O'$.

\subsection{Graph canonization and Orbits ordering} 
In graph theory, graph canonization is the problem of finding a canonical form of a given graph $G$. A canonical form is a labeled graph $\Canon{G}$ that is isomorphic to $G$, and such that every graph that is isomorphic to $G$ has the same canonical form as $G$. Thus, from a solution to the graph canonization problem, one could also solve the problem of graph isomorphism: to test whether two graphs $G'$ and $G''$ are isomorphic, compute their canonical forms $\Canon{G'}$ and $\Canon{G''}$, and test whether these two canonical forms are identical. The formal computational study of graph canonization began in the 1970s. Quite recently, in~\cite{Babai19}, L{\'{a}}szl{\'{o}} Babai announced a quasi-polynomial-time algorithm for graph canonization, that is, one with running time $2^{O((\log n)^{c})}$ for some fixed $c>0$ and $n$ being the number of vertices. Available algorithms can compute not only the canonical labeling but also the orbits.\footnote{e.g. \emph{bliss}, see \url{http://www.tcs.tkk.fi/Software/bliss/index.html}} 

\begin{figure}[t]
  \centering
  \resizebox{0.8\textwidth}{!}{%
       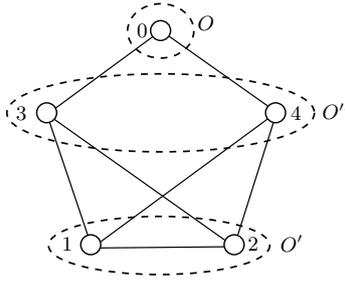    
       }
       \caption{An example of graph canonization. Note that $O < O' < O''$, according to the vertex labeling provided by the canonization.}
          \label{fig:canon}
\end{figure}

Consider the situation where two robots, during the same \Look phase, take a snapshot of the graph $G$. Since they are disoriented, each robot may represent $G$ differently (say as $G'$ and $G''$). However, if they both apply the same canonization algorithm, they obtain $\Canon{G'}$ and $\Canon{G''}$, which must be identical since both derive from the same underlying graph $G$. In particular, if the canonization assigns integer labels from $[0,n-1]$ to the vertices of any $n$-vertex graph, then the vertex labeled $i$ in $\Canon{G'}$ is isomorphic to the vertex labeled $i$ in $\Canon{G''}$. Consequently, the corresponding orbits in $\orbs{O}_{G'}$ and $\orbs{O}_{G''}$ receive the same set of labels. This ensures that the two robots can agree on a total ordering of the orbits of $G$: $O'< O''$ if the smallest label assigned to $O'$ is smaller than the smallest label assigned to $O''$ (cf. Fig.~\ref{fig:canon}).

% ------------------------------
\subsection{Our approach and Methodology} 
The algorithms introduced in this paper to solve the {\gath} problem on general non-vertex-transitive graphs follow the methodology proposed in \cite{CDN21a}. We briefly outline how a generic algorithm $\mathcal{A}$, intended to solve a problem $\mathcal{P}$, can be designed according to this approach.

Since robots have extremely limited capabilities, it is  beneficial to decompose $\mathcal{P}$ into simpler %sub-problems. Each sub-problem is associated with a ``task'' that can be executed by one or more robots. Let us denote these 
tasks 
denoted 
as $T_1, T_2, \dots, T_q$. Among these tasks, at least one is designated as the \emph{terminal} one.% , that is, the main problem $\mathcal{P}$ has been solved and no further action is required.

%Since we operate within a model where robots have extremely limited capabilities, it is usually helpful to decompose the main problem $\mathcal{P}$ into simpler sub-problems. Each sub-problem is associated with a ``task'' that can be executed by one or more robots. Let us denote these tasks as $T_1, T_2, \dots, T_q$. At least one of such tasks is referred to as the \emph{terminal} one. It represents the case in which $\mathcal{P}$ has been solved and hence robots just have to recognize this and perform no moves. 

Since robots operate according to the {\tt LCM} cycle, they must identify the correct task to execute based on the configuration they observe during the {\tt Look} phase. This task recognition is triggered by associating a predicate $P_i$ with each task $T_i$. %When a robot wakes up and detects that a predicate $P_i$ holds, it understands that it must execute task $T_i$.
More concretely, if the predicates are well-defined, %algorithm $\mathcal{A}$ can use them during the {\tt Compute} phase as follows: if 
a robot $r$ observing that predicate $P_i$ is true, can executes the corresponding move $m_i$ associated with task $T_i$.
To ensure the correctness of this method, each predicate must satisfy the following properties:

\begin{itemize}
    
    \item $Prop_1$: each predicate $P_i$ must be computable based on the configuration $C$ observed during the {\tt Look} phase; 
    
    \item$Prop_2$: the predicates must be mutually exclusive, i.e., $P_i \wedge P_j = \texttt{false}$ for all $i \neq j$, ensuring that each robot unambiguously selects a single task;
    
    \item$Prop_3$: for every possible configuration $C$, there must be at least one predicate $P_i$ that is evaluated as true.
    
\end{itemize}

To define each predicate $P_i$, we first identify a set of basic variables that capture relevant properties of the configuration $C$, e.g., metric, numerical, ordinal, or topological features, that can be computed by each robot using only its local observation. Then, such variables are used to assemble a pre-condition $\texttt{pre}_i$ that must be satisfied for task $T_i$ to be applicable. Finally, predicate $P_i$ can be defined as:
\[P_i = \texttt{pre}_i \wedge \neg(\texttt{pre}_{i+1} \vee \texttt{pre}_{i+2} \vee \dots \vee \texttt{pre}_q).\]
This definition guarantees that $Prop_2$ is always satisfied and imposes a specific order on the task evaluation. Specifically, predicates are evaluated in reverse order: the robot first checks $P_q = \texttt{pre}_q$, then $P_{q-1} = \texttt{pre}_{q-1} \wedge \neg \texttt{pre}_q$, and so on. If all predicates from $P_q$ down to $P_2$ evaluate to false, then $P_1$ must be true and task $T_1$ executed.
When a robot performs a generic task $T_i$ in a configuration $C$, the algorithm may lead to a new configuration $C’$ where another task $T_j$ must be performed. In such a case, we say that the algorihtm induces a transition from $T_i$ to $T_j$. Collectively, all such transitions form a directed graph called the \emph{transition graph}. The terminal task, which marks the successful resolution of problem $\mathcal{P}$, must correspond to a sink vertex in this graph.
As shown in~\cite{CDN21a}, the correctness of an algorithm designed in this way is guaranteed if the following conditions hold:
\begin{description}
    \item[$H_1$:] the transition graph is correct, i.e., for each task $T_i$, the reachable tasks are exactly those depicted in the transition graph;
    \item[$H_2$:] all the loops in the transition graph, including self-loops not involving sink vertices, must be executed a finite number of times;
    \item[$H_3$:] with respect to the studied problem $\mathcal{P}$, no a-priori proved unsolvable configuration is generated by 
        $\mathcal{A}$.
\end{description}

% ------------------------------------
% NON VERTEX-TRANSITIVE
% ------------------------------------
\section{Algorithm for graphs with terminal orbits}\label{sec:non-vertex}

In this section, we consider configurations defined on connected and non-vertex-transitive graphs. In other words, any graph $G$ considered here to define a configuration admits at least two orbits. We restrict here the analysis to non-vertex-transitive graphs that admit \emph{terminal orbits}, as introduced in the following definition. 

\begin{definition}\label{def:terminal}
Let $G$ be a non-vertex-transitive graph and $O$ be an orbit of $G$. We say that $O$ is \emph{terminal} if the following property holds:
\begin{equation}
\forall  u\in V(G)\setminus O,~ \forall v\in O,~ \exists u,v\textrm{-path } P \textrm{ such that } P\cap O = \{v\}.
\end{equation} 
\end{definition}

\begin{figure}[t]
  \centering
  \resizebox{0.8\textwidth}{!}{%
    \begin{minipage}{\textwidth}
      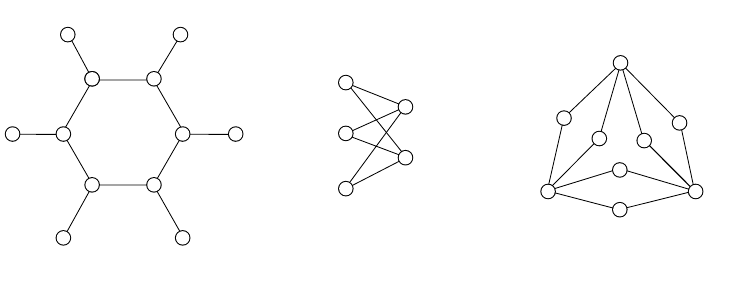
          \end{minipage}%
          }
          \caption{Three graphs with two orbits each. The first one admits one terminal orbit constituted by the pendant vertices. In the second one, that is a complete bipartite graph $K_{3,2}$, the two orbits are constituted by the two partitions, resp., and they are both terminal. In the last graph, one orbit is constituted by the vertices of degree four and the other by the vertices of degree two, and they are both not terminal (cf. Def.~\ref{def:terminal}).}
          \label{fig:terminal}
\end{figure}

Informally, when an orbit $O$ is terminal in $G$, it is possible to move from any vertex $u$ not in $O$ to any vertex $v$ in $O$ without traversing vertices in $O$ except when reaching $v$. This is equivalent to saying that removing all but one vertex of $O$ produces a connected induced subgraph of $G$.
Fig.~\ref{fig:terminal} shows graphs admitting or not terminal orbits. In the first case, the orbit containing all the pendant vertices is terminal (but the second orbit containing all the remaining vertices is not terminal). In the second example, each side of the complete bipartite graph is a terminal orbit. In the latter example, there are two orbits (one of them formed by the vertices with degree 4), and it is easy to check that none of them is terminal.

% \begin{figure}[h]
%     \centering    \includegraphics[width=0.4\linewidth]{terminal.jpeg}
% \includegraphics[width=0.20\linewidth]{no-terminal.jpeg}
%     \caption{The first two graphs admit terminal orbits; the last one does not contain any terminal orbit.}
%     \label{fig:terminal}
% \end{figure}

Let $G$ be a non-vertex-transitive graph, $O$ be its smallest terminal orbit, and $C = (G, \lambda)$ an initial configuration. We define a gathering algorithm $\algoterm$ based on just three tasks: %. For each task, we list the corresponding precondition and move:
\begin{description}
\item[$T_1$:]     
$O$ has two or more occupied vertices. If the active robot $r$ is in $O$, then $r$ moves out of $O$;
\item[$T_2$:] 
$O$ has one occupied vertex $v$. If 
the active robot $r$ is in $O$, then $r$ does not move. If the active robot $r$ is outside $O$, then $r$ moves toward $v$ without passing through $O$ before reaching $v$;
\item[$T_3$:] 
$O$ has no occupied vertices. The active robot $r$ moves toward an arbitrary vertex $v$ in $O$.
\end{description}

%The following statement shows that a simple gathering algorithm can be provided for configurations based on graphs admitting terminal orbits. 

\begin{theorem}\label{teo:terminal}
Given a non-vertex-transitive graph $G$ and an initial configuration $C=(G,\lambda)$, if $G$ has terminal orbits, algorithm $\algoterm$ is correct under \rr, requiring $\Theta(diam(G))$ epochs.
\end{theorem}
\begin{proof}

We prove that this algorithm solves the gathering problem in $O(diam(G))$ epochs. If at time $t_1=0$ task $T_1$ is executed in $C$, since the smallest orbit $O$ is always computable by all robots, the algorithm proceeds by keeping $T_1$ under execution until time $t_2 > t_1$ when the condition that activates $T_2$ holds. This happens after at most $k-1$ robots move out of $O$. The whole task then requires at most one epoch. Each robot takes one \LCM\ cycle to do so because $G$ is connected and every vertex in $O$ must admit an edge leading outside $O$. From time $t_2$, the configuration evolves according to one or more executions of $T_2$ until time $t_3>t_2$, when the gathering will be achieved. This holds because $O$ is terminal, so during the robots' movements toward orbit $O$, this orbit always remains with only one occupied vertex. In this case, each robot may perform at most $O(diam(G))$ moves, hence the whole process may require at most $O(diam(G))$ epochs. The same analysis holds if $T_2$ (instead of $T_1$) is performed at time $t_1=0$. Conversely, if $T_3$ is performed at time $t_1=0$, as soon as the first robot reaches $O$ within one epoch, $T_2$ is performed in the obtained configuration, and therefore gathering is assured. Overall, $\algoterm$ requires $O(diam(G))$ epochs to solve {\gath}. According to Lemma~\ref{lem:lb}, $\algoterm$ is basically time optimal as the initial configuration $C$ might admit $\Delta(C)=diam(G)$.\qed
\end{proof}

The following statement provides a property that can be used to characterize graphs not admitting terminal orbits.

\begin{theorem}\label{teo:connected-termional}
Given a non-vertex-transitive graph $G$, if it admits a proper subset of orbits whose vertices induce a connected subgraph of $G$, then $G$ has a terminal orbit. 
\end{theorem}
\begin{proof}
Let $\orbs{O}_G$ be the set containing all the orbits of $G$, and let $\orbs{O}'$ be any proper subset of $\orbs{O}_G$ whose vertices induce a connected subgraph $G'$ of $G$.
Let  $O\in \orbs{O}_G\setminus \orbs{O}'$ %$P$ 
%be another orbit of $G$ not belonging to $\mathcal{O}$ but 
an orbit with vertices adjacent to vertices in $G'$.
%that admits connections with $G'$. 
By definition of orbit, any vertex in $O$ %P$ 
%admits a connection with $G'$. 
is adjacent to a vertex in $G'$.
Since $G'$ is connected, we have that any vertex $v\in O$ %$P$ 
can be reached from any vertex in $u\in G'\setminus O$ through a path $P$ such that $P\cap O = \{v\}$ , i.e. $O$ %$P$ 
is terminal with respect to the graph $G''$ induced by the vertices in $\orbs{O}'\cup \{O\}$. Furthermore, we notice that $G''$ 
%the vertices in $\orbs{O}'\cup \{O\}$ %$\mathcal{O}\cup \{P\}$ 
is a connected subgraph of $G$ larger than $G'$.
Hence, any orbit $O'$ %$Q$ 
not belonging to $\orbs{O}'\cup \{O\}$ %$\mathcal{O}\cup \{P\}$ 
but adjacent to them is terminal with respect to the subgraph induced by the vertices in $\orbs{O}'\cup \{O,O'\}$. %$\mathcal{O}\cup \{P,Q\}$.
If we keep proceeding in this way, since $G$ is connected, then we conclude that there must exist at least one orbit that is terminal with respect to the entire graph $G$.\qed
\end{proof}

Notice that the opposite is not true; in fact, in the complete bipartite graph $K_{m,n}$ with $m\neq n$, there are two orbits, each of which is terminal, but neither orbit induces a connected subgraph. Moreover, the previous theorem can be alternatively expressed as follows: if $G$ is a non-vertex-transitive graph that does not contain any terminal orbit, then every proper subset of orbits of $G$ induces a disconnected subgraph. In particular, \emph{each orbit of $G$ must induce a disconnected subgraph}.

% ------------
\section{Graph classes and terminal orbits}
This section aims to show examples of non-vertex-transitive graphs that admit or do not admit terminal orbits.

% ----------------
\subsection{Graphs with terminal orbits}
We first establish relationships between typical graph properties and terminal orbits, and then we use such properties to analyze some graph classes. To this end, we need to recall some additional basic graph-theoretic terminology. The \emph{eccentricity} of a vertex $v$ is the maximum distance between $v$ and any other vertex of the graph. The \emph{center} of a graph $G$, denoted $c(G)$, is the set of vertices of $G$ with minimum eccentricity. A \emph{block} of $G$ is any maximal subgraph of $G$ that contains no cut-vertex. If $G$ is connected, it is known that $c(G)$ is contained within some block of $G$ (e.g., see~\cite[Theorem 12.5]{Chartrand2012}). 

\begin{theorem}\label{teo:graph_properties}
Let $G$ be a non-vertex-transitive graph. $G$ admits a terminal orbit if one of the following properties holds: $G$ has a universal vertex, $G$ has a cut-vertex, $G$ has an orbit whose vertices are pairwise false twins or pairwise true twins.
\end{theorem}
\begin{proof}
In the whole proof, let $G$ be any non-vertex-transitive graph. 

Assume that $G$ has a universal vertex $v$ (i.e., $v$ is adjacent to all other vertices of the graph). If $O_v$ is the orbit of $G$ containing $v$, it directly follows from Def.~\ref{def:terminal} that $O_v$ is a terminal orbit of $G$. 

Consider now the case in which $G$ has a cut-vertex. 
%Let $B$ be the block of $G$ containing $c(G)$, and $v_1,v_2,\ldots,v_k$, $k\ge 1$, be the cut vertices contained in $B$. Removing $v_i$ disconnects $G$ into two connected components, one containing $B$ and the other denoted as $V_i$. If $k>1$ we have components $V_1,V_2,\ldots,V_k$; the case $k=1$ may generate just $V_1$ or $V_1$ and $V_2$  (the latter holds when the unique cut-vertex is the only vertex in $c(G)$). Now, denote as $v'_1$ any vertex in $V_1$ at largest distance from $v_1$, and as $O_{v'_1}$ the orbit of $G$ containing $v'_1$. Notice that $O_{v'_1}$ may have several vertices in each component $V_i$. We claim that $O_{v'_1}$ is terminal. To show that, we prove in the following that for each $u\not\in O_{v'_1}$ and for each $v\in O_{v'_1}$, there exists a $u,v$-path $P$ such that $P\cap O_{v'_1} = \{v'_1\}$. Without loss of generality, assume that $v\equiv v'_1$. 
%If $u\in V_1$, then $P$ starts with a shortest $u,v_1$-path followed by a shortest $v_1,v'_1$-path. 
%If $u\in V_i$, $i\ge 1$, then $P$ starts with a shortest $u,v_i$-path followed by a shortest $v_i,v'_1$-path. If $u\in B$, then $P$ is any shortest $u,v_i$-path followed by a shortest $v_i,v'_1$-path. In any case, according to the definition of $v'_1$, we easily get that each vertex of $P$, except $v'_1$, does not belong to $O_{v'_1}$.
Let $B$ be a block of $G$ containing all the vertices in $c(G)$.  Consider the connected components obtained from $G$ by removing  a cut vertex $u$ of $B$. Let $C$ one of this components such that $V(C)\cap V(B)=\emptyset$. Let $v\in V(C)$ be a vertex at maximum distance from $u$. Let $O_v$ be the orbit to which $v$ belongs.
We claim that $O_v$ is terminal.
To show that, we prove in the following that for each $x\not\in O_{v}$ and for each $y\in O_{v}$, there exists a $x,y$-path $P$ such that $V(P)\cap O_{v} = \{y\}$. Without loss of generality, assume that $y\equiv v$.  

If $x\in V(G)\setminus V(C)$, then $P$ starts with a shortest $x,u$-path followed by a shortest $u,v$-path. By definition of $v$, we easily find that each vertex of $P$, except $v$, does not belong to $O_{v}$.
If $x\in V(C)$, consider the subgraph $H$ induced a shortest $x,u$-path and a shortest $u,v$-path. By definition of $v$ and $x$, we have that $V(H)\cap O_v=\{v\}$, then there exists a path $P$ in $H$ such that $V(P)\cap O_v=\{v\}$.

Finally, assume that $G$ has an orbit $O$ whose vertices are pairwise twins. If the vertices in $O$ are all true twins, then $O$ induces a connected subgraph. In this case, Theorem~\ref{teo:connected-termional} guarantees that $G$ has a terminal orbit. If $O$ contains all false twins, then we get $N(v)=N(v')$ for each $v,v'\in O$. This implies that for each $u\in V(G)\setminus O$ there exists shortest a $u,v$-path $P$ that does not pass through any $v'$ twin of $v$, i.e., $P\cap O = \{v\}$. Hence, $O$ is a terminal orbit of $G$.
\qed
\end{proof}

This theorem can now be used to show that graphs belonging to well-known graph classes admit terminal orbits.
A \emph{cactus graph} is any graph in which every block is an edge or a cordless cycle. A \emph{block graph} is any graph in which every block is a complete graph. A \emph{trivially perfect graph} is a graph with the property that in each of its induced subgraphs, the size of the maximum independent set equals the number of maximal cliques. A \emph{threshold graph} is a graph that can be constructed from a one-vertex graph by repeated applications of the following two operations: (1) addition of a single isolated vertex to the graph, and (2) addition of a single vertex that is connected to all other vertices. A \emph{windmill graph} $Wd(m,n)$ is an undirected graph constructed for $m\ge 2$ and $n\ge 2$ by joining $n$ copies of the complete graph $K_m$ at a shared universal vertex. 
%The corona product of graphs $G$ and $H$, denoted $G\circ H$, can be obtained by taking one copy of $G$, called the center graph, and a number of copies of $H$ equal to the order of $G$. Then, each copy of $H$ is assigned a vertex in $G$, and that one vertex is attached to each vertex in its corresponding $H$ copy by an edge.
With $t G$ we denote the graph consisting in $t$ copies of a graph $G$, whereas with $G_1 + G_2$ we denote the \emph{join} of graphs $G_1$ and $G_2$, that is the graph given by a copy of $G_1$, a copy of $G_2$, and all the edges joining $V(G_1)$ and $V(G_2)$.

\begin{corollary}\label{cor:graph_classes}
Let $G$ be a non-vertex-transitive graph belonging to any of the following graph classes:  
cactus graphs, 
block graphs, 
wheel graphs,
trivially perfect graphs,
threshold graphs,
windmill graphs, 
 $G' + t K_1$, and $G' + K_t$, where $G'$ is any arbitrary non-complete connected graph.
Then, $G$ admits a terminal orbit. 
\end{corollary}
\begin{proof}
Let $G$ be a cactus or a block graph. Since $G$ is a non-vertex-transitive it is clear that $G$ has a cut-vertex. 

If $G$ is a trivially perfect graph, a threshold graph, or a windmill graph, then $G$ contains a universal vertex (e.g., see~\cite{BLS99}). 

%Finally, let $G$ be a corona graph $G = G'\circ K_1$, where $G'$ is any arbitrary connected graph with $n$ vertices. Denote as $O$ the subset of vertices of $G$ corresponding to the  $n$ copies of $K_1$. Since each vertex in $O$ is connected to each vertex of $G'$, it follows that $O$ is an orbit of $G$ formed by pairwise false twins.  

Finally, in  $G=G' + t K_1$, denote as $O$ the subset of vertices of $G$ corresponding to the $t$ copies of $K_1$. Since each vertex in $O$ is connected to each vertex of $G'$, and $G'$ is not complete, it follows that $O$ is an orbit of $G$ formed by pairwise false twins.  Similarly, in $G=G' + K_t$, the vertices of $K_t$ form an orbit of pairwise true twins.

In all the above cases, Theorem~\ref{teo:graph_properties} implies that $G$ has a terminal orbit.
\qed
\end{proof}

% ----------------
\subsection{Graphs without terminal orbits}
We show here that there are graphs not admitting terminal orbits, even among well-known graph classes.

Let $\mathcal{H}_{A,B}$ be a family containing infinitely many graphs denoted as $G(A,B;n)$ and defined as follows. Select two arbitrary but distinct vertex-transitive graphs $A$ and $B$, an integer $n\ge 3$, and build $G(A,B;n)$ by fist taking $n$ copies $A_0,A_1,\ldots, A_{n-1}$ of $A$ and $n$ copies $B_0,B_1,\ldots, B_{n-1}$ of $B$, and then making the joins $A_i + B_i$ and $B_i + A_{i+1}$,%
\footnote{Here, the addition is intended modulo $n$. 
%The join $G=G_1 + G_2$ of graphs $G_1$ and $G_2$ with disjoint point sets $V_1$ and $V_2$ is the graph union $G_1\cup G_2$ together with all the edges joining $V_1$ and $V_2$. 
}
for each $i$, $0\le i\le n-1$. As an example, the last graph in Fig.~\ref{fig:terminal} represents the graph $G(2K_1,K_1;3)$.

\begin{figure}[t]
    \centering
    \includegraphics[width=0.6\linewidth]{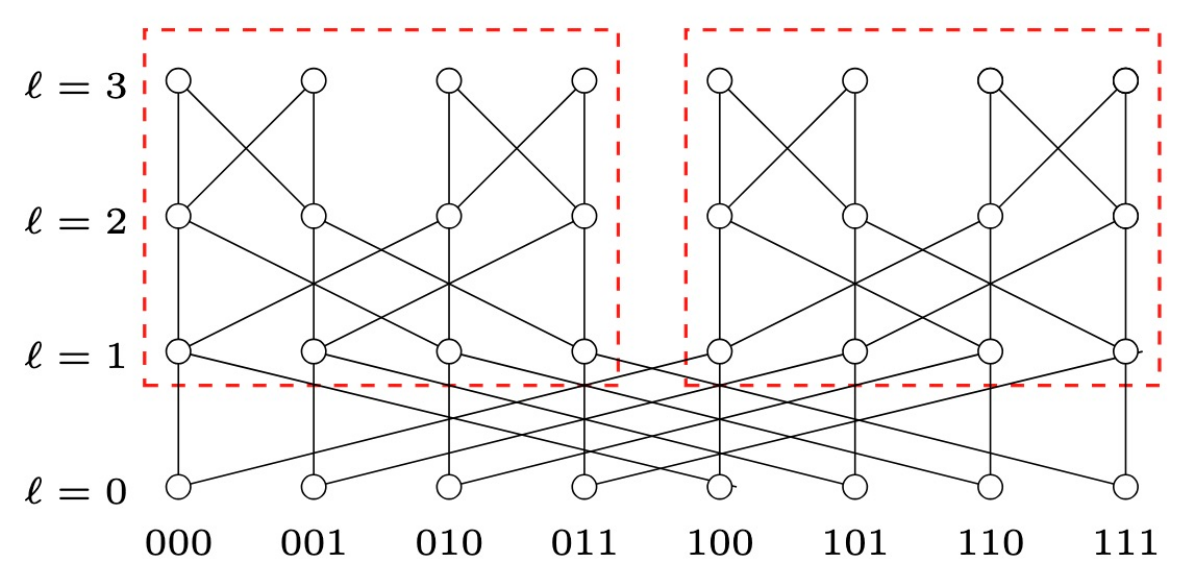}
    \caption{A visualization of $\BF(3)$. The dashed rectangles contain two copies of $\BF(2)$ that, along with additional vertices for layer 0, form $\BF(3)$.}    
    \label{fig:butterfly}
\end{figure}

Before showing that graphs in $\mathcal{H}_{A,B}$ do not admit a terminal orbit, we also remind another well-known class of graphs, the $d$-dimensional butterflies. A $d$-dimensional butterfly $\BF(d)$, $d\ge 1$, is an undirected graph with vertices $[\ell,c]$, where $\ell\in \{0,1,\ldots,d\}$ is the \emph{layer} and $c\in \{0,1\}^d$ is the \emph{column}. The vertices $[\ell,c]$ and $[\ell',c']$ are adjacent if $|\ell - \ell'| = 1$, and either $c = c'$ or $c$ and $c'$ differ precisely in the $\ell$-th bit.  The vertices at layer $0$ and $d$ have degree $2$, whereas the rest of the vertices have degree $4$. We call layer $0$ and $d$ \emph{boundary layers}. It is easy to observe that each orbit $O_i$ of $\BF(d)$ is composed of \emph{complementary} layers $i$ and $d-i$, with $0\le i \le \dmezzi $. By \emph{intermediate orbit} we mean $O_{d/2}$ when $d$ is even. Notice that the intermediate orbit is formed by one layer only, while the other orbits are formed by two layers each; moreover, $O_0$ corresponds to the union of the boundary layers.
Note that $\BF(d)$ can also be recursively built by using two copies of $\BF(d-1)$ along with $2^d$ additional vertices, those forming layer 0. As an example, see Fig.~\ref{fig:butterfly}.

\begin{theorem}
If $G$ belongs to $\mathcal{H}_{A,B}$ or $G$ is a butterfly graph, then $G$ does not admit a terminal orbit.
\end{theorem}
\begin{proof}
%\ser{per la dimostrazione bisogna per forza usare la definizione, ovvero mostrare che ogni orbita è non terminale}
Consider an arbitrary graph $G(A,B;n)$. It can be easily observed that such a graph has two orbits: $O_A$ formed by the union of the $n$ copies of $A$ and $O_B$ formed by the union of the $n$ copies of $B$. Now, if $u\in B_i$ and $v\in O_A\setminus \{A_i,A_{i+1}\}$, then the ``circular structure'' of the graph implies that each $u,v$-path must necessarily include a vertex belonging to a copy of $B$. This proves that neither $O_A$ nor $O_B$ is terminal.

Let $B(d)$ be a butterfly graph of order $d\ge 1$. If $d=1$, then the butterfly coincides with a cycle $C_4$, a graph that cannot have terminal orbits because vertex-transitive. For $d\ge 2$, denote as $\BF'$ and $\BF''$ the two copies of the $\BF(d-1)$ that, along with vertices at layer 0, form $\BF(d)$ (cf.  Fig.~\ref{fig:butterfly}). 
We now prove that each orbit $O_i$ is not terminal. If $i=0$, then consider $v$ at layer $d$ but in $\BF'$ and $u\in O_j\cap \BF''$, $j>0$. It is clear that each $u,v$-path contains a vertex at layer 0, that is, a vertex of $O_0$ different from $v$. This implies that $O_0$ is not a terminal orbit.
If $i>0$, then consider $v\in O_i$, in particular $v$ is located at layer $d-i$ of $\BF'$, and $u\in O_j$, in particular $u$ is located in a layer $k>d-i$ of $\BF''$. It is clear that each $u,v$-path contains a vertex at layer $d-i$ of $\BF''$, that is a vertex on $O_i$ different from $v$. Hence, also in this case $O_i$ is not a terminal orbit.\qed
\end{proof}

%A function $\lambda: V\to \mathcal{N}$ provides the number of robots located on a given vertex of $G$. We call $C=(G,\lambda)$ a \emph{configuration} whenever the total number of robots located on the vertices of $G$ is bounded and greater than zero. A function $\isOcc: V\to \{true, false\}$ simply indicates whether a given vertex is occupied by one or more robots or is empty (i.e., no robots are on that vertex). Given a configuration $C=(G,\lambda)$, we call $C^p=(G,\isOcc)$ the \emph{perceived configuration} of $C$ -- i.e., the configuration $C$ from the robots' perspective.

% ---------------------------
%vecchia sezione Algorithm for graphs without terminal orbits spostata in
%\input{Algorithm_NoTerminalOrbit_old.tex}
% \input{Algorithm_NoTerminalOrbit_notsoold.tex}
\section{Algorithm for graphs without terminal orbits}
In this section, we describe $\algonoterm$, an algorithm able to solve the {\gath} problem in each configuration $C=(G,\lambda)$ defined on a connected non-vertex-transitive graph $G$ \emph{without terminal orbits}. It is worth remarking that, according to Theorem~\ref{teo:connected-termional}, each orbit $O$ of $G$ induces a disconnected subgraph $G[O]$. Moreover, each connected component of $G[O]$ must be vertex-transitive, with vertices adjacent to the vertices of another orbit, since the whole graph $G$ is connected.

% -------------------
\subsection{Algorithm description} 
The strategy of $\algonoterm$ is to confine all robots within a subgraph $G'$ of $G$ that admits a terminal orbit. Hence, applying $\algoterm$ for graphs admitting terminal orbits on the configuration restricted to $G'$ would solve the gathering (Task $T_4$). 
Subgraph $G'$ is constructed as follows. 
At each activation, during the {\tt Compute} phase, a robot applies the graph canonization method in order to obtain an ordered sequence of the orbits of $G$. Let $O_1$ be the smallest orbit and $O_2$ be the first orbit in the sequence that is adjacent to $O_1$. The subgraph $G'$ is composed of one connected component $CC$ of $G[O_2]$, along with all vertices in $O_1$ that are adjacent to any vertex in $CC$. 
Ideally, in a first phase (Task $T_1$), all robots are moved to the vertices of $O_1$.
In the second phase (Task $T_{2.i}$), one robot moves from a vertex in $O_1$ to a vertex in $O_2$, defining the connected component $CC$. 
In the third phase (Tasks $T_{2.ii}$, $T_{2.iii}$), all the remaining robots should move from $O_1$ to $CC$. 

Naturally, the paths of these robots are not always direct from $O_1$ to $O_2$; they may pass through other orbits (even all orbits may be involved). To manage this, our algorithm allows at most one connected component of $O_2$ to contain occupied vertices, and permits exactly one additional robot to move through the graph to reach $CC$. However, if the moving robot $r$ is part of a multiplicity (i.e., multiple robots sharing a vertex), all robots in that multiplicity are moved one by one together with $r$ (Task $T_{2.iv}$).

Since there are cases where $CC$ contains just one occupied vertex and the moving robot steps over $O_2$ as well, there might be some `confusion' in detecting $CC$, hence we have a fourth phase (Task $T_{3.i}$) that deals with such special occurrences. Also in this phase, if the moving robot $r$ is on a multiplicity, all the robots in the multiplicity are moved with $r$ one by one (Tasks $T_{3.ii}$ and $T_{3.iii}$).

In what follows, we formalize each of these tasks. Note that, according to the methodology sketched in Section~\ref{sec:model}, we assume that the proposed algorithm $\algonoterm$ evaluates the preconditions that must hold for each task in the reverse order from $pre_4$ to $pre_1$, and performs the first task for which the precondition is true. 

\begin{description}
\item[Task 1:] 
If none of the preconditions $pre_4$, $pre_3$, and $pre_2$ hold for a given configuration, algorithm $\algonoterm$ makes all the robots that are not on a vertex of $O_1$ move toward $O_1$. 

\noindent {\bf Precondition $pre_1$}: $true$.

\noindent {\bf Move $m_1$}: any robot not in $O_1$ moves toward $O_1$.

\item[Task 2:] 
This task is the core of algorithm $\algonoterm$. If all robots are in $O_1$, any robot in $O_1$ moves toward $O_2$. If all robots are in $O_1$ or in a single connected component of $G[O_2]$, called $CC$, 
any robot on a vertex of $O_1$ closest to $CC$ moves toward $CC$.
If there is a single occupied vertex $u$ except those in $CC$ and $O_1$, which is also closer to $CC$ than any other occupied vertex in $O_1$, any robot on $u$ moves toward $CC$. 
Unfortunately, it could be the case that on $u$ there is a multiplicity, then, after a move from $u$ to say $v$, both vertices $u$ and $v$ are occupied and the algorithm deals with this case by moving all the remaining robots on $u$ toward $v$ before continuing.

\noindent{\bf Precondition $pre_2$}: the set of occupied vertices $V'$ is the union of three sets $F$, $U$ and $M$ such that:
\begin{itemize}
\item $F$ is the set of all occupied vertices in $O_1$, $|F|\ge 0$;
\item $U$ is the set of all occupied vertices in a connected component $CC$ in $G[O_2]$;
\item $M$ consists of an occupied vertex $u$, if any, closer to $CC$ than any occupied vertex in $F$ and an occupied vertex $u$, if present, such that  $uv \in E$  and $v$ is on a shortest path from $u$ to $CC$. 
In any case, $M\cup F$ and $U$ form a partition of $V'$.
\end{itemize}

\noindent Furthermore, %the implication $|M|>0\implies |U|>0$ holds, together with 
one of the following conditions holds:
\begin{description}
\item[i)] $|M|=0$, $|U|=0$, $|F|>0$;
\item[ii)] $|M|=0$, $|U|>0$, $|F|>0$;
\item[iii)] $M=\{u\}$, $|U|>0$, $|F|\ge 0$;
\item[iv)] $M=\{u,v\}$, $|U|>0$, $|F|\ge 0$.
\end{description}

\noindent{\bf Move $m_2$}:\\
if i), any robot in $O_1$ moves toward $O_2$;\\
if ii), any robot on a vertex of $F$ and closest to $CC$ moves toward $CC$;\\
if iii), any robot on $u$ moves toward $CC$;\\
if iv), any robot on $u$ moves toward $v$.

\item[Task 3:] 
This task deals with all the configurations in which the connected component $CC$ of $G[O_2]$, where the robots are supposed to go, could be confused with another component of $G[O_2]$. If this happens, it means there are exactly two connected component in $G[O_2]$ containing robots. We call $CC_1$ and $CC_2$ these two components.

\noindent{\bf Precondition $pre_3$}: the set of occupied vertices is the union of three sets $F$, $U_1$, and $U_2$ such that:
\begin{itemize}
\item $F$ is the set of all occupied vertices in $O_1$, $|F|\ge 0$;
\item $U_1$ ($U_2$, resp.) consists of an occupied vertex $v$ in $CC_1$ ($CC_2$, resp.) 
closer to $CC_2$ ($CC_1$, resp.) than any occupied vertex in $F$ and an occupied vertex $u$, if present, such that  $uv \in E$  and $v$ is on a shortest path from $u$ to $CC_2$ ($CC_1$, resp.). 
\end{itemize}

\noindent Furthermore, one of the following conditions holds:
\begin{description}
\item[i)] $|U_1|=1$, $|U_2|=1$;
\item[ii)] $|U_1|=1$, $|U_2|=2$;
\item[iii)] $|U_1|=2$, $|U_2|=2$.
\end{description}

\noindent {\bf Move $m_3$}:\\
if i), any active robot in $U_1$ ($U_2$, resp.) moves toward $CC_2$ ($CC_1$, resp.);\\
if ii), any active robot on $u$ moves toward vertex $v$;\\
if iii), any active robot on $u$ in $U_1$ ($U_2$, resp.) moves toward vertex $v$.\\

\item[Task 4:] 
This task deals with a configuration where all the robots are on a subgraph $G'$ of $G$ having a terminal orbit, that is gatherable by applying $\algoterm$, cf. Theorem~\ref{teo:terminal}.

\noindent {\bf Precondition $pre_4$}: all  robots are on vertices of a subgraph $G'$ composed by one connected
component $CC$ belonging to $G[O_2]$ and all the vertices in $O_1$ adjacent to vertices in $CC$.

\noindent {\bf Move $m_4$}: apply the algorithm designed for terminal configurations confined to $G'$.
\end{description}

\subsection{Correctness}
With the next lemmas, we prove that condition $H_1$ holds, that is, the correctness of the transition graph described by Table~\ref{tab:transgraph}. Then, with Theorem~\ref{teo:algo}, we also prove that condition $H_2$ holds. With respect to condition $H_3$, this is satisfied as there are no forbidden configurations.

\begin{table}[t]
    \centering
    \begin{tabular}{c|l}
    \hline
$T_1$  & $T_2$, $T_3$, $T_4$\\\hline
$T_{2.i}$ & $T_{2.ii}$  \\
$T_{2.ii}$ & $T_{2.ii}$, $T_{2.iii}$, $T_{2.iv}$,  $T_3$\\
$T_{2.iii}$ & $T_{2.ii}$, $T_{2.iii}$, $T_{2.iv}$, $T_3$, $T_4$\\
$T_{2.iv}$ & $T_{2.iii}$, $T_{2.iv}$, $T_{3.i}$, $T_{3.ii}$, $T_4$\\\hline
$T_{3.i}$ & $T_{2.iii}$, $T_{2.iv}$,  $T_{3.i}$, $T_{3.ii}$, $T_4$\\
$T_{3.ii}$ & $T_{3.i}$, $T_{3.ii}$\\
$T_{3.iii}$ & $T_{3.ii}$, $T_{3.iii}$\\\hline
$T_4$ & {\gath}\\\hline
    \end{tabular}\newline 
    \caption{A tabular representation of the transition graph of the algorithm for non-vertex-transitive graphs without terminal orbits.}
    \label{tab:transgraph}
\end{table}

\begin{lemma}\label{lem:t4}
    Let $C$ be a configuration in $T_1$. From $C$, $\algonoterm$ leads to a configuration belonging to $T_2$, $T_3$, or $T_4$.
\end{lemma}

\begin{proof}
According to move $m_1$, any robot that awakes outside $O_1$ moves toward $O_1$. As long as the configuration remains in $T_1$, robots have to move at most $diam(G)$ hops, hence requiring at most $diam(G)$ epochs. If at some point the obtained configuration does not belong to $T_1$ anymore, it may concern any other task among $T_2$, $T_3$, or $T_4$. Notice that we are guaranteed that such a transition occurs because if not earlier, it happens once all the robots have reached $O_1$ and $pre_2$ with condition i) holds.\qed
\end{proof}

\begin{lemma}\label{lem:t3}
    Given a configuration  $C$ in $T_2$, $\algonoterm$ from $C$ leads to a configuration belonging to $T_{2.ii}$, $T_{2.iii}$, $T_{2.iv}$, $T_3$ or $T_4$.
\end{lemma}

\begin{proof}
According to move $m_2$, there are four possible cases.
When $|U|>0$, let $G'$ be the subgraph of $G$ consisting of the connected component $CC$ belonging to $O_2$ and all vertices in $O_1$ adjacent to vertices in $CC$.

\begin{itemize}
\item[i)] When $|M|=0$, $|U|=0$ and $|F|>0$, then the first robot that awakes moves toward $O_2$, and the configuration is necessarily in $T_{2.ii}$,  as $|M|=0$, $|U|=1$ and $|F|>0$. The configuration achieved cannot be in $T_4$ since that would imply that $C$ was already in $T_4$. The configuration achieved cannot be in $T_3$ since there is just one robot out of $O_1$.
\item[ii)] When $|M|=0$, $|U|>0$ and $|F|>0$ then any robot $r$ in $O_1$, closest to the only connected component $CC$ occupied by robots in $O_2$, moves toward it. The move may lead to a configuration $C'$ still in $T_{2.ii}$, if $r$ moves on a vertex of $CC$. 
$C'$ might be in $T_{2.iii}$ or $T_{2.iv}$ (if $r$ moves from a multiplicity) or, similarly, in $T_3$ if the robot moves on a vertex of $O_2$.

$C'$ cannot be in $T_4$ since there is at least another robot at the same original distance of $r$ from $CC$, otherwise $|M|>0$.

\item[iii)] When $M=\{u\}$, $|U|>0$ $|F|\ge 0$, then any robot on $u$ moves toward $CC$. If there is only one robot on $u$, the resulting configuration could remain in $T_{2.iii}$ or lead to a configuration in $T_{2.ii}$, if the robot reaches a vertex of $CC$. The resulting configuration could be in $T_3$ if the robot reaches a vertex in $O_2$, or even in $T_4$, if the moving robot reaches a vertex in $G'$. If, instead, there is a multiplicity in $u$, then the movement may lead to $T_{2.iv}$  with $|M|=2$, $|U|>0$ and $|F|\ge 0$.

\item[iv)] When $|M|=\{u,v\}$, $|U|>0$ $|F|\ge 0$,  any robot on $u$ moves on $v$ toward $CC$. If there is a multiplicity on $u$, the resulting configuration clearly remains in $T_{3.iv}$. If there is a single robot on $u$, the resulting configuration could be in  $T_{2.iii}$, in $T_4$ (if $v$ is in $O_1$ and adjacent to a vertex in $CC$) or even in $T_{3.i}$  or $T_{3.ii}$ (if $v$ is in $O_2$ and $u$ on a different orbit).\qed
\end{itemize}
\end{proof}

\begin{lemma}\label{lem:t2}
Given a configuration  $C$ in $T_3$, $\algonoterm$ from $C$
    leads to a configuration belonging to  $T_{2.ii}$, $T_{2.iii}$, $T_{2.iv}$, $T_3$ or $T_4$.
\end{lemma}

\begin{proof}
In case $C$ is in $T_{3.i}$, according to move $m_{3.i}$ one robot on a vertex in $U_1\cup U_2$ moves. Without loss of generality, assume that the active robot is on $u$ and moves toward $CC_2$. Assume first no multiplicity is on $u$. Of course, in one step a moving robot cannot reach its target since by definition there are no edges between $CC_1$ and $CC_2$. If the reached vertex $w$ is in $O_2$ the obtained configuration still belongs to $T_{3.i}$, otherwise the configuration is in $T_{2.iii}$, or even in $T_4$ if $u$ is in $G'$.  
If there is a multiplicity in $u$, the reached configuration still has occupied vertices in two connected components of $G[O_2]$. Then the configuration is in $T_{3.ii}$, or in $T_{2.iv}$ with $M=\{u,w\}$ and $|U|=1$.

In case $C$ is in $T_{3.ii}$, assume $v$ be the vertex closest to $CC_1$ and $U_1=\{u,v\}$. Any active robot on $u$  moves to $v$. If there is a multiplicity in $u$, the resulting configuration is clearly still in $T_{3.ii}$. Otherwise, the configuration is in $T_{3.i}$.

In case $C$ is in $T_{3.iii}$, any active robot on $u$ of either $U_1$ or $U_2$ moves to $v$. If there is a multiplicity in $u$, the resulting configuration is clearly still in $T_{3.iii}$, otherwise it is in $T_{3.ii}$.\qed
\end{proof}

\begin{lemma}\label{lem:t1}
Given a configuration  $C=(G,\lambda)$ in $T_4$, $\algonoterm$ from $C$  solves {\gath} in at most $O(|V(G)|)$ epochs.
\end{lemma}
\begin{proof}
Let $G'$ be the subgraph of $G$ induced by the unique connected component $CC$ occupied by robots in $O_2$ along with each neighboring vertex in $O_1$. It is easy to observe that $G'$ can be considered non-vertex-transitive by exploiting the distinction among vertices in $O_1$ and $O_2$. Moreover, there are vertices in $O_1$ that belong to a terminal orbit of $G'$. By considering the sub-configuration $C'=(G',\lambda)$, according to Theorem~\ref{teo:connected-termional} we can use algorithm $\algonoterm$ to solve the  {\gath} problem in $G'$ (all robots are gathered in a vertex belonging to $O_1$). 
By Theorem~\ref{teo:terminal}, the algorithm requires $O(diam(G'))$ epochs, which in turn is bounded by $O(|V(G)|)$.\qed
\end{proof}

\begin{theorem}\label{teo:algo}
Given a non-vertex-transitive graph $G$ with no terminal orbits
and an initial configuration $C=(G,\lambda)$, the {\gath} problem can be solved form $C$ under the \rr\ scheduler in $O(occ(C)\cdot \Delta(C) + |V(G)| )$ epochs.
\end{theorem}

\begin{proof}
By Table~\ref{tab:transgraph}, clearly Task $T_1$ can be performed only from an initial configuration since no transition leads to it. By move $m_1$, each robot is moved on a vertex of $O_1$ tracing a path of length at most $\Delta(C)$. Then, the task is performed within $\Delta(C)$ epochs.
Similarly, Task $T_{2.i}$, where all robots are in $O_1$ and one of them moves to $O_2$, is performed at most once. This task requires only one move.
Furthermore, also Task $T_4$ is performed at most once. By Lemma~\ref{lem:t1}, it requires at most $|V(G)|$ epochs. The overall number of epochs required by the above tasks is then bounded by $O(|V(G)|)$.

It remains to analyze the other cases, that is, configurations in $T_{2.ii}$, $T_{2.iii}$, $T_{2.iv}$ or $T_3$. Note that, if from one of the tasks $T_{2.ii}$, $T_{2.iii}$ and $T_{3.i}$, a robot $r$ moves from a multiplicity and the achieved configuration is in one of the tasks among $T_{2.iv}$, $T_{3.ii}$ and $T_{3.iii}$, then, by the respective moves $m_{3.ii}$, $m_{3.iii}$ and $m_{2.iv}$, all the robots in the multiplicity reach $r$ within one epoch. 
So, in what follows, we consider only configurations in $T_{2.ii}$, $T_{2.iii}$ and $T_{3.i}$, without considering the transient configurations in $T_{2.iv}$, $T_{3.ii}$ and $T_{3.iii}$. On the one hand, it allows us to analyze the movement of a multiplicity like a move of a single robot. On the other hand, we assume that moving a robot to an adjacent vertex requires one epoch, since it could be the case it moved from a multiplicity.

%To analyze the remaining cases,
%let $V'$ be the set of occupied vertices in a configuration, and  $\ell=|V'|$. 
We now show that the total number $B$ of robots in at most two components of $G[O_2]$ always increases.

Let us analyze the case in which no configuration in $T_3$ is generated, that is, a connected component $CC$ of $G[O_2]$, possibly generated after tasks $T_1$ or $T_{2.i}$, or present in the initial configuration, is always recognizable.

Assume that the configuration is $T_{2.ii}$, that is, all occupied vertices are in $O_1$ or in $CC$. Then, a robot $r$ on a vertex $v$ closest to $CC$ moves toward $CC$. From now on, the generated configuration is in $T_{2.iii}$, until a configuration in $T_3$ or $T_4$ is generated or the moving robot reaches $CC$.  In conclusion, in a configuration of tasks $T_{2.ii}$ and $T_{2.iii}$, a moving robot $r$ and all robots in the same multiplicity of $r$,  reach a configuration in $T_{2.ii}$, $T_3$ or $T_4$ in $diam(G)$ epochs. Meanwhile, the number $B$ of robots in $CC$ (or in $CC_1$ and $CC_2$ for a new configuration in $T_3$) increases by at least one.

It remains to discuss the case in which the configuration is in $T_3$ that is, it is not possible to distinguish a single connected component in $G[O_2]$. Recall that, if $C$ is in $T_{3.ii}$ or in $T_{3.iii}$, the repeated application of the corresponding moves eventually generates a configuration in $T_{3.i}$ within one epoch. Meanwhile, the connected components $CC_1$ and $CC_2$ do not change.  

Then, consider a configuration $C$ in $T_{3.i}$ and let $B$ be the number  of robots in $CC_1$ and $CC_2$. Without loss of generality, let $r_1$ be the activated robot in $CC_1$ that moves toward $CC_2$. The obtained configuration is then in $T_{2.iii}$ as soon as $r_1$ moves outside $CC_1$. Now, from the above discussion, a configuration in $T_4$ is generated, or again a configuration in $T_{3.i}$ with the occupied vertex in $CC_1$ closer to $CC_2$ then in $C$. 
Symmetrically, let $r_2$ be the robot in $CC_2$ eventually activated and moving toward $CC_1$. Then, either the two robots meet in a single component $CC$ or a configuration in $T_4$ is generated. This requires at most $\Delta(C)$ epochs. The total number of robots $B$ in $CC_1$ and $CC_2$ is not increased, but the obtained configuration, if not in $T_4$, is in now in $T_{2.ii}$ or $T_{2.iii}$ and from there, $B$
will increase again.

In summary, moving the robots from a vertex occupied by $t\geq 1$ robots from a configuration in $T_2$ or $T_3$ to a new configuration in $T_2$ or $T_4$ requires $O(\Delta(C))$ epochs, and if the resulting configuration is in $T_2$, value $B$ increases by $t$.
%
%\ser{Ricontrollare la frase che segue, non chiara. Inoltre, mettere $k$ al posto di $n$?} 
Since value $B$ cannot be greater than $k$ (i.e., it cannot increase more than $k$), the algorithm converges to a configuration in $T_4$, eventually.
The total number of epochs required for moving all robots from the initial occupied vertices of configurations in $T_2$ and $T_3$ is then bounded by $O(occ(C)\cdot \Delta(C))$.

In conclusion, the overall number of epochs to achieve the gathering is $O(occ(C)\cdot \Delta(C) + |V(G)| )$.
\qed
\end{proof}

% ------------------------------------
% CONCLUSION
% ------------------------------------
\section{Concluding remarks and future work}\label{sec:concl}
We have approached the {\gath} problem for a swarm of robots moving on non-vertex-transitive graphs under a Round Robin scheduler.

We have designed two general resolution algorithms dealing with configurations admitting or not terminal orbits, respectively. In particular, we have proposed a  time-optimal algorithm, $\algoterm$, designed for graphs that contain a terminal orbit. For graphs that lack terminal orbits, we have designed an algorithm, $\algonoterm$, that for any initial configuration $C=(G,\lambda)$ guarantees the {\gath} within $O(occ(C)\cdot \Delta(C) + |V(G)|)$ epochs. 
It remains open whether it is possible to provide a time optimal resolution algorithm with respect to the basic lower bound of $\Omega(\Delta(C)$ provided by Lemma~\ref{lem:lb}.

As a research direction for future work, we aim to investigate the {\gath} problem in vertex-transitive graphs.

% ------------------------------------
% BIBLIOGRAPHY
% ------------------------------------
%\bibliographystyle{splncs04}
%\bibliography{global_references,local}

\end{document}

%% file: commands.tex
\definecolor{ser}{rgb}{0.95, 0.1, 0.1}
\definecolor{alf}{HTML}{0030f3}
\definecolor{ale}{HTML}{1d8348}
\definecolor{gab}{HTML}{dd13F8}

\newcommand{\Look}{{\tt Look}\xspace}

\newcommand{\LCM}{{\tt LCM}\xspace}

\newcommand{\A}{\mathcal{A}} % generic algorithm
\newcommand{\Ex}{\mathbb{E}} 
\newcommand{\rr}{$\mathsf{RR}$}
\newcommand{\Aut}[1]{\mbox{Aut}({#1})}

\newcommand{\Canon}[1]{\mbox{Canon}({#1})}
\newcommand{\orbs}[1]{\mathcal{#1}}

\newcommand{\algoterm}{\mathcal{A}_{\mathbf{T}}}
\newcommand{\algonoterm}{\mathcal{A}_{\neg\mathbf{T}}}

\newcommand{\BF}{\mathit{BF}}
\newcommand{\dmezzi}{\lfloor d/2 \rfloor}

\newcommand{\block}[1]{\smallskip\noindent{\textbf{#1.}}~}

\newcommand{\gath}{\sc{Gathering}\xspace}
\newcommand{\dgath}{\sc{Distinct Gathering}\xspace}

\SetKwComment{tcp}{$\#$ }{}%

\SetCommentSty{mycommfont}

%% file: canon.pdf_tex
%% Accompanies image file 'canon.pdf' (pdf, eps, ps)
\begingroup%
  \makeatletter%
  \providecommand\color[2][]{%
    \errmessage{(Inkscape) Color is used for the text in Inkscape, but the package 'color.sty' is not loaded}%
    \renewcommand\color[2][]{}%
  }%
  \providecommand\transparent[1]{%
    \errmessage{(Inkscape) Transparency is used (non-zero) for the text in Inkscape, but the package 'transparent.sty' is not loaded}%
    \renewcommand\transparent[1]{}%
  }%
  \providecommand\rotatebox[2]{#2}%
  \newcommand*\fsize{\dimexpr\f@size pt\relax}%
  \newcommand*\lineheight[1]{\fontsize{\fsize}{#1\fsize}\selectfont}%
  \ifx\svgwidth\undefined%
    \setlength{\unitlength}{375.92135752bp}%
    \ifx\svgscale\undefined%
      \relax%
    \else%
      \setlength{\unitlength}{\unitlength * \real{\svgscale}}%
    \fi%
  \else%
    \setlength{\unitlength}{\svgwidth}%
  \fi%
  \global\let\svgwidth\undefined%
  \global\let\svgscale\undefined%
  \makeatother%
  \begin{picture}(1,0.33932378)%
    \lineheight{1}%
    \setlength\tabcolsep{0pt}%
    \put(0,0){\includegraphics[width=\unitlength,page=1]{canon.pdf}}%
    \put(0.53243388,0.29472887){\color[rgb]{0,0,0}\makebox(0,0)[lt]{\lineheight{1.25}\smash{\begin{tabular}[t]{l}$O$\end{tabular}}}}%
    \put(0.67615225,0.19142746){\color[rgb]{0,0,0}\makebox(0,0)[lt]{\lineheight{1.25}\smash{\begin{tabular}[t]{l}$O''$\end{tabular}}}}%
    \put(0.62782773,0.03829534){\color[rgb]{0,0,0}\makebox(0,0)[lt]{\lineheight{1.25}\smash{\begin{tabular}[t]{l}$O'$\end{tabular}}}}%
    \put(0.4627713,0.28735468){\color[rgb]{0,0,0}\makebox(0,0)[lt]{\lineheight{1.25}\smash{\begin{tabular}[t]{l}$0$\end{tabular}}}}%
    \put(0.37603819,0.03945638){\color[rgb]{0,0,0}\makebox(0,0)[lt]{\lineheight{1.25}\smash{\begin{tabular}[t]{l}$1$\end{tabular}}}}%
    \put(0.59042344,0.03945638){\color[rgb]{0,0,0}\makebox(0,0)[lt]{\lineheight{1.25}\smash{\begin{tabular}[t]{l}$2$\end{tabular}}}}%
    \put(0.32332051,0.19058035){\color[rgb]{0,0,0}\makebox(0,0)[lt]{\lineheight{1.25}\smash{\begin{tabular}[t]{l}$3$\end{tabular}}}}%
    \put(0.63962655,0.19058035){\color[rgb]{0,0,0}\makebox(0,0)[lt]{\lineheight{1.25}\smash{\begin{tabular}[t]{l}$4$\end{tabular}}}}%
  \end{picture}%
\endgroup%

%% file: nonVertex.pdf_tex
%% Accompanies image file 'nonVertex.pdf' (pdf, eps, ps)
\begingroup%
  \makeatletter%
  \providecommand\color[2][]{%
    \errmessage{(Inkscape) Color is used for the text in Inkscape, but the package 'color.sty' is not loaded}%
    \renewcommand\color[2][]{}%
  }%
  \providecommand\transparent[1]{%
    \errmessage{(Inkscape) Transparency is used (non-zero) for the text in Inkscape, but the package 'transparent.sty' is not loaded}%
    \renewcommand\transparent[1]{}%
  }%
  \providecommand\rotatebox[2]{#2}%
  \newcommand*\fsize{\dimexpr\f@size pt\relax}%
  \newcommand*\lineheight[1]{\fontsize{\fsize}{#1\fsize}\selectfont}%
  \ifx\svgwidth\undefined%
    \setlength{\unitlength}{361.38820798bp}%
    \ifx\svgscale\undefined%
      \relax%
    \else%
      \setlength{\unitlength}{\unitlength * \real{\svgscale}}%
    \fi%
  \else%
    \setlength{\unitlength}{\svgwidth}%
  \fi%
  \global\let\svgwidth\undefined%
  \global\let\svgscale\undefined%
  \makeatother%
  \begin{picture}(1,0.37900264)%
    \lineheight{1}%
    \setlength\tabcolsep{0pt}%
    \put(0,0){\includegraphics[width=\unitlength,page=1]{nonVertex.pdf}}%
  \end{picture}%
\endgroup%